\documentstyle[12pt,axodraw]{article}

\catcode`@=12
\begin{document}
\thispagestyle{empty}
\begin{flushright} 
UCRHEP-T297\\ 
TPI-MINN-01/01\\
UMN-TH-1933\\
January 2001\
\end{flushright}
\vspace{0.5in}
\begin{center}
{\Large	\bf Realistic Gluino Axion Model Consistent with \\ Supersymmetry 
Breaking at the TeV Scale\\}
\vspace{1.5in}
{\bf D. A. Demir$^1$ and Ernest Ma$^2$\\}
\vspace{0.2in}
{\sl $^1$ Theoretical Physics Institute, University of Minnesota, Minneapolis, 
MN 55455\\}
{\sl $^2$ Physics Department, University of California, Riverside, 
California 92521\\}
\vspace{1.5in}
\end{center}
\begin{abstract}\
The recently proposed model of using the dynamical phase of the gluino to 
solve the strong CP problem is shown to admit a specific realization in 
terms of fundamental singlet superfields, such that the breaking of 
supersymmetry occurs only at the TeV scale, despite the large axion scale 
of $10^9$ to $10^{12}$ GeV.  Phenomenological implications are discussed.
\end{abstract}
\newpage
\baselineskip 24pt

The strong CP problem is the problem of having the instanton-induced term 
\cite{cpv}
\begin{equation}
{\cal L}_\theta = \theta_{QCD} {g_s^2 \over 64 \pi^2} \epsilon_{\mu \nu 
\alpha \beta} G_a^{\mu \nu} G_a^{\alpha \beta}
\end{equation}
in the effective Lagrangian of quantum chromodynamics (QCD), where $g_s$ 
is the strong coupling constant, and
\begin{equation}
G_a^{\mu \nu} = \partial^\mu G_a^\nu - \partial^\nu G_a^\mu + g_s f_{abc} 
G_b^\mu G_c^\nu
\end{equation}
is the gluonic field strength.  If $\theta_{QCD}$ is of order unity, the 
neutron electric dipole moment is expected \cite{edm} to be $10^{10}$ 
times its present experimental upper limit ($0.63 \times 10^{-25}~e$ cm) 
\cite{nedm}.  This conundrum is most elegantly resolved by invoking 
a dynamical mechanism \cite{pq} to relax the above $\theta_{QCD}$ parameter 
(including all contributions from colored fermions) to zero.  However, this 
necessarily results \cite{ww} in a very light pseudoscalar particle called 
the axion, which has not yet been observed \cite{search}.

To reconcile the nonobservation of an axion in present experiments and the 
constraints from astrophysics and cosmology \cite{astro}, two types of 
``invisible'' axions are widely discussed.  The DFSZ solution \cite{dfsz} 
introduces a heavy singlet scalar field as the source of the axion but 
its mixing with the doublet scalar fields (which couple to the usual quarks) 
is very much suppressed.  The KSVZ solution \cite{ksvz} also has a heavy 
singlet scalar field but it couples only to new heavy colored fermions.

Consider now the incorporation of supersymmetry into the Standard Model (SM) 
of particle interactions.  The list of colored fermions consists of not 
only the usual quarks, but also the gluinos.  The parameter $\theta_{QCD}$ 
of Eq.~(1) is then replaced by
\begin{equation}
\overline \theta = \theta_{QCD} - Arg~Det~M_u M_d - 3~Arg~M_{\tilde g},
\end{equation}
where $M_u$ and $M_d$ are the respective mass matrices of the charge 2/3 and 
$-1/3$ quarks, and $M_{\tilde g}$ is the gluino mass.  Our recent proposal 
\cite{dema,demasa} is to relax $\overline \theta$ to zero with the dynamical 
phase of the gluino, instead of the quarks as in the DFSZ model or other 
unknown colored fermions as in the KSVZ model.  The source of this axion 
is again a heavy singlet scalar field, but since its vacuum expectation 
value (VEV) is supposed to be in the range $10^9$ to $10^{12}$ GeV to 
satisfy the astrophysical and cosmological bounds \cite{astro}, 
supersymmetry is expected to be broken at that scale as well, and not 
at 1 TeV as desired.  This is of course also a problem in the 
supersymmetric versions of the DFSZ and KSVZ models.  In the following 
we will show how it gets resolved in a specific 
realization of the gluino axion model \cite{dema,demasa}.

Our {\it first} key observation is the identification of the anomalous global 
symmetry $U(1)_R$ of supersymmetric transformations as the $U(1)_{PQ}$ 
symmetry which solves the strong CP problem and generates the axion. 
Under $U(1)_R$, the scalar components of a chiral superfield transform 
as $\phi \to e^{i \theta R} \phi$, whereas the fermionic components 
transform as $\psi \to e^{i \theta (R-1)} \psi$.  In the minimal 
supersymmetric standard model (MSSM), the quark and lepton superfields 
$\hat Q$, $\hat u^c$, $\hat d^c$, $\hat L$, $\hat e^c$ have $R=+1$, 
whereas the Higgs superfields $\hat H_u$, $\hat H_d$ have $R=0$.  The 
superpotential
\begin{equation}
\hat W = \mu \hat H_u \hat H_d + h_u \hat H_u \hat Q \hat u^c + h_d \hat H_d 
\hat Q \hat d^c + h_e \hat H_d \hat L \hat e^c
\end{equation}
has $R=+2$ except for the $\mu$ term (which has $R=0$).  Hence the resulting 
Lagrangian breaks $U(1)_R$ explicitly, leaving only a discrete remnant, i.e. 
the usual $R$ parity: $R = (-1)^{3B+L+2J}$.  The gluino axion model 
\cite{dema,demasa} replaces $\mu$ with a singlet superfield of $R=+2$ 
and requires the entire theory to be invariant under $U(1)_R$, which is 
then spontaneously broken.  The reason that $U(1)_R$ is a natural choice 
for $U(1)_{PQ}$ is that the gauginos of the MSSM have $R=+1$, hence 
the phase of the gluino mass must be dynamical and contributes to 
$\overline \theta$ of Eq.~(3).  In fact, all SM particles have $R=0$ 
(i.e. even $R$ parity) and all superparticles have $R = \pm 1$ (i.e. odd 
$R$ parity), but the only colored fermions with $R \neq 0$ are the gluinos. 
In the minimal SM with only one Higgs doublet, there is no $U(1)_{PQ}$, 
hence both the DFSZ and KSVZ models require additional particles.  In the 
MSSM, there is also no $U(1)_{PQ}$, but if the $\mu$ term and the soft 
supersymmetry breaking $A$ terms and gaugino masses are removed, then the 
$U(1)_R$ symmetry is available for us to identify as the $U(1)_{PQ}$ symmetry 
for solving the strong CP problem.  Of course, we still need to implement 
this idea with a specific choice of additional particles.  However, no 
matter what we do, we are faced with a fundamental problem (which also 
exists if we want to consider supersymmetric versions of the DFSZ and 
KSVZ models): if the superfield containing the axion is spontaneously 
broken at $10^9$ to $10^{12}$ GeV, how is the supersymmetry preserved 
down to the order of 1 TeV?

Our {\it second} key observation has to do with the consequence of the 
spontaneous breaking of a global symmetry without breaking the 
supersymmetry.  Because the supersymmetry is not broken, there has to be 
a massless superfield, the scalar component of which is {\it complex}. 
In addition to the usual phase degree of freedom, there is now also 
a {\it scale} degree of freedom, hence such models always contain an 
indeterminate mass scale \cite{spont}.  The trick then is to construct 
a realistic gluino axion model using this ambiguity of scale so that 
the subsequent soft breaking of supersymmetry occurs at 1 TeV, but the 
vacuum expectation value of the scalar field containing the axion is 
$10^9$ to $10^{12}$ GeV.

Following Ref.\cite{demasa}, we introduce again three singlet superfields 
$\hat S_2$, $\hat S_1$, and $\hat S_0$, with $R = 2, 1, 0,$ respectively 
and impose the $Z_3$ discrete symmetry under which $\hat S_1$ and $\hat S_0$ 
transform as $\omega$ and $\hat S_2$ as $\omega^2$, with $\omega^3 = 1$. 
The most general superpotential with $R=2$ containing these superfields 
is then given by
\begin{equation}
\hat W = m_2 \hat S_2 \hat S_0 + f_1 \hat S_1 \hat S_1 \hat S_0.
\end{equation}
Let $\hat S_{2,1,0}$ be replaced by $v_{2,1,0} + \hat S_{2,1,0}$, then
\begin{eqnarray}
\hat W &=& m_2 v_0 \hat S_2 + 2 f_1 v_1 v_0 \hat S_1 + (m_2 v_2 + f_1 v_1^2) 
\hat S_0 \nonumber \\ 
&+& f_1 v_0 \hat S_1 \hat S_1 + (m_2 \hat S_2 + 2 f_1 v_1 \hat S_1) \hat S_0 
+ f_1 \hat S_1 \hat S_1 \hat S_0.
\end{eqnarray}
Hence the minimum of the corresponding scalar potential is given by
\begin{equation}
V_{min} = |m_2 v_0|^2 + 4|f_1 v_1 v_0|^2 + |m_2 v_2 + f_1 v_1^2|^2.
\end{equation}
To preserve supersymmetry, we need $V_{min} = 0$.  Hence
\begin{equation}
v_0 = 0, ~~~ v_2 = - {f_1 v_1^2 \over m_2}.
\end{equation}
If $v_{1,2} \neq 0$, then $U(1)_R$ is spontaneously broken, but the scale 
of symmetry breaking is indeterminate \cite{spont} because only the ratio 
$v_1^2/v_2$ is constrained.  Moreover, since $m_2$ is presumably very large, 
say of the order of some unfication scale,
\begin{equation}
v_2 << v_1
\end{equation}
is predicted, unless of course both $v_1$ and $v_2$ are of order $m_2$. 
(Exactly what value each actually takes will depend on the subsequent 
soft breaking of the supersymmetry as we will show later.)  With this 
solution,
\begin{equation}
\hat W = {m_2 \over v_1} (v_1 \hat S_2 - 2 v_2 \hat S_1) \hat S_0 
+ f_1 \hat S_1 \hat S_1 \hat S_0,
\end{equation}
which shows clearly that the linear combination
\begin{equation}
{v_1 \hat S_1 + 2 v_2 \hat S_2 \over \sqrt {|v_1|^2 + 4|v_2|^2}}
\end{equation}
is a massless superfield.  Hence the axion is mostly contained in $S_1$, 
but since only $S_2$ couples to the MSSM particles, the effective axion 
coupling to gluinos is $(v_2/v_1) v_2^{-1} = v_1^{-1}$ as desired.

Consider now the breaking of the supersymmetry by soft terms \underline 
{at the TeV scale} which preserve the $U(1)_R$ symmetry but are allowed to 
break the $Z_3$ discrete symmetry \cite{whepp2}.  We start with the original 
superpotential of Eq.~(5), write down its corresponding scalar potential, and 
add all such soft terms regardless of whether or not they are holomorphic, 
i.e.
\begin{eqnarray}
V &=& |m_2 S_0|^2 + 4|f_1 S_0 S_1|^2 + |m_2 S_2 + f_1 S_1^2|^2 \nonumber \\ 
&+& \mu_0^2 |S_0|^2 + \mu_1^2 |S_1|^2 + \mu_2^2 |S_2|^2 \nonumber \\ 
&+& [\mu_{12} S_1^2 S_2^* + \mu_{00} S_0 |S_0|^2 + \mu_{01} S_0 |S_1|^2 
+ \mu_{02} S_0 |S_2|^2 + h.c.]
\end{eqnarray}
The minimum of $V$ is now determined by
\begin{eqnarray}
&& v_0 \simeq - {\mu_{01} v_1^2 \over m_2^2}, \\ 
&& v_1^2 \simeq {\mu_1^2 m_2 \over 4 f_1 \mu_{12}} \left[ 1 - {\mu_{12} \over 
2 f_1 m_2} + {f_1 \mu_2^2 \over 2 m_2 \mu_{12}} + {\mu_2^2 \over m_2^2} 
\right], \\ 
&& v_2 \simeq - {\mu_1^2 \over 4 \mu_{12}} \left[ 1 - {\mu_{12} \over f_1 m_2} 
+ {f_1 \mu_2^2 \over 2 m_2 \mu_{12}} \right].
\end{eqnarray}
Hence $v_0 << v_2 << v_1$ and the supersymmetric solution of Eq.~(8) remains 
valid to a very good approximation.  (In Ref.\cite{demasa}, $\mu_{12}$ was 
written as $\lambda m_2$ with the implicit assumption that it is of order 
$m_2$, hence $\mu_1$ in that case is of order $v_1$.)  We now realize the 
important fact that all soft supersymmetry breaking parameters ($\mu_1, 
\mu_{12}, etc.$) can be of order 1 TeV so that $v_2$ is of order 1 TeV 
as shown by Eq.~(15), and yet $v_1$ is larger than $v_2$ 
by a factor of order $\sqrt {m_2/v_2}$.

Consider now the physical masses of $S_{2,1,0}$ and their fermionic partners. 
The linear combination given by Eq.~(11) still contains the axion, but 
because supersymmetry (as well as $Z_3$) is broken at the TeV scale, its 
fermionic component (axino) is allowed to have a Majorana mass of that 
magnitude. The phase of its bosonic component is the axion, but the magnitude 
(maxion) is a scalar field of mass $\sqrt {-2 \mu_1^2}$ (instead of zero).  
The orthogonal combination to that given by Eq.~(11) combines with $\hat S_0$ 
to form a heavy superfield containing two complex scalars and a Dirac fermion 
of mass $m_2$ as expected.  Hence the low-energy particle content of our model 
consists of ($i$) a Majorana fermion at the TeV scale, ($ii$) a real scalar 
field also at the TeV scale, and ($iii$) an axion which couples only to 
superparticles. Since all the above particles come from mostly $\hat S_1$ but 
only $\hat S_2$ interacts directly with the MSSM particles, their effects 
are generally suppressed by the factor $v_2/v_1$.

Our {\it third} key observation has to do with how gauginos acquire mass in 
the presence of $U(1)_R$.  In Refs.\cite{dema,demasa}, the explicit arbitrary 
supersymmetry-breaking term $S_2^* \tilde g \tilde g$ is assumed. 
Since this is not a soft term, it is not clear how it can be justified 
rigorously.  Here we show that it is actually generated through loop 
corrections.  The so-called $A$ terms are also generated through their 
effective couplings to $S_2$.  Both gaugino masses and $A$ terms are 
forbidden by $U(1)_R$ invariance, but are allowed as $v_1$ and $v_2$ 
become nonzero.

The MSSM superpotential of Eq.~(4) is now replaced by
\begin{eqnarray}
\hat W &=& h_2 \hat S_2 \hat H_u \hat H_d + h_u \hat H_u \hat Q \hat u^c + 
h_d \hat H_d \hat Q \hat d^c + h_e \hat H_d \hat L \hat e^c \nonumber \\ 
&+& m_2 \hat S_2 \hat S_0 + f_1 \hat S_1 \hat S_1 \hat S_0,
\end{eqnarray}
where we have assumed that $\hat H_u, \hat H_d$ transform as $\omega^2$, 
$\hat Q, \hat L$ as $\omega$ and $\hat u^c, \hat d^c, \hat e^c$ as 1 
under $Z_3$.  The breaking of supersymmetry is achieved by the soft terms 
of Eq.~(12) together with
\begin{eqnarray}
V_{soft} &=& \tilde Q^\dagger M^2_Q \tilde Q + \tilde u^{c \dagger} M^2_{u^c} 
\tilde u^c + \tilde d^{c \dagger} M^2_{d^c} \tilde d^c + \tilde L^\dagger 
M^2_L \tilde L + \tilde e^{c \dagger} M^2_{e^c} \tilde e^c \nonumber \\ 
&+& M^2_{H_u} |H_u|^2 + M^2_{H_d} |H_d|^2 + (M^2_{ud} H_u H_d + h.c.)
\end{eqnarray}
This differs from that of the MSSM only in that the $A$ terms and the gaugino 
masses are absent because they are not invariant under $U(1)_R$.  The usual 
$\mu B$ term of the MSSM is denoted as $M^2_{ud}$ here because the parameter 
$\mu$ is now absent.

From Eq.~(16), we find the following terms,
\begin{equation}
|h_2 S_2 H_d + h_u \tilde Q \tilde u^c|^2 + |m_2 S_0 + h_2 H_u H_d|^2,
\end{equation}
in the Lagrangian of our model.  Together with the $H_u H_d$ term in Eq.~(17), 
we obtain an effective interaction (see Fig.~1, left window) given by
\begin{equation}
\left(M^2_{ud} + {h_2 \over f_1} v_2 \mu_{01} \right) {h_2^* h_u \over 
M^2_{H_d}} S_2^* H_u \tilde Q \tilde u^c,
\end{equation}
where Eqs.~(13) to (15) have been used. This means that an effective $A$ 
term is generated as $S_2$ is replaced by its VEV, i.e.
\begin{equation}
A_u = {h_2^* h_u v_2 \over M^2_{H_d}} \left(M^2_{ud} + {h_2 \over f_1} v_2 
\mu_{01} \right),
\end{equation}
which has the desirable feature of being proportional to $h_u$ and thus the 
automatic suppression of flavor-changing neutral currents from the 
supersymmetric scalar sector.

\begin{figure}[h]
\centerline{
{\epsfysize = 5cm \epsffile{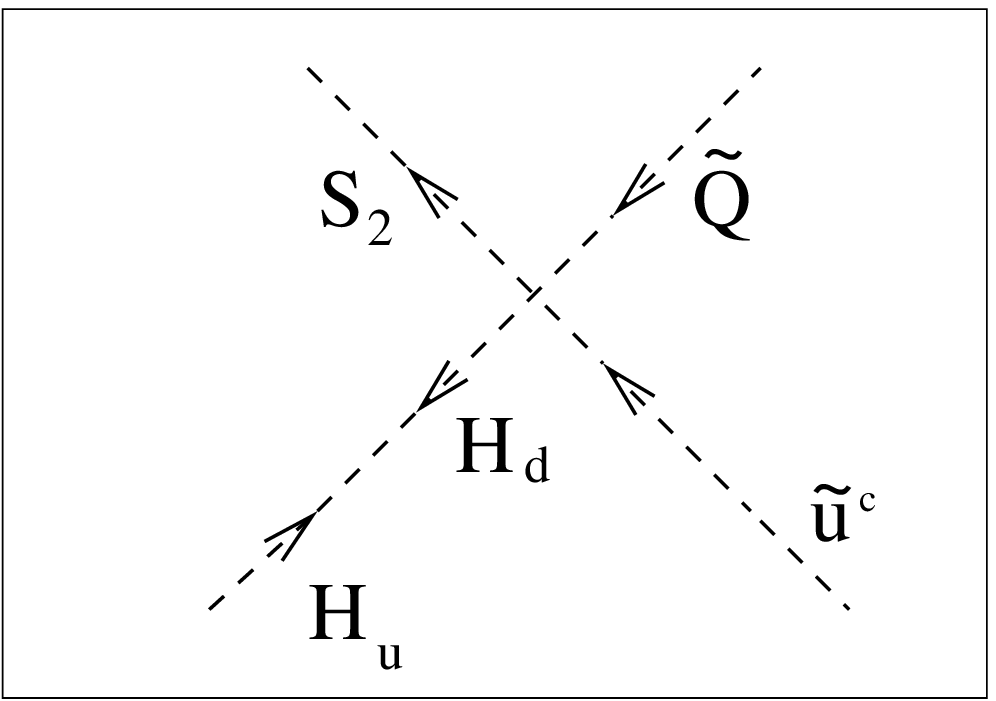}} {\epsfysize = 5cm \epsffile{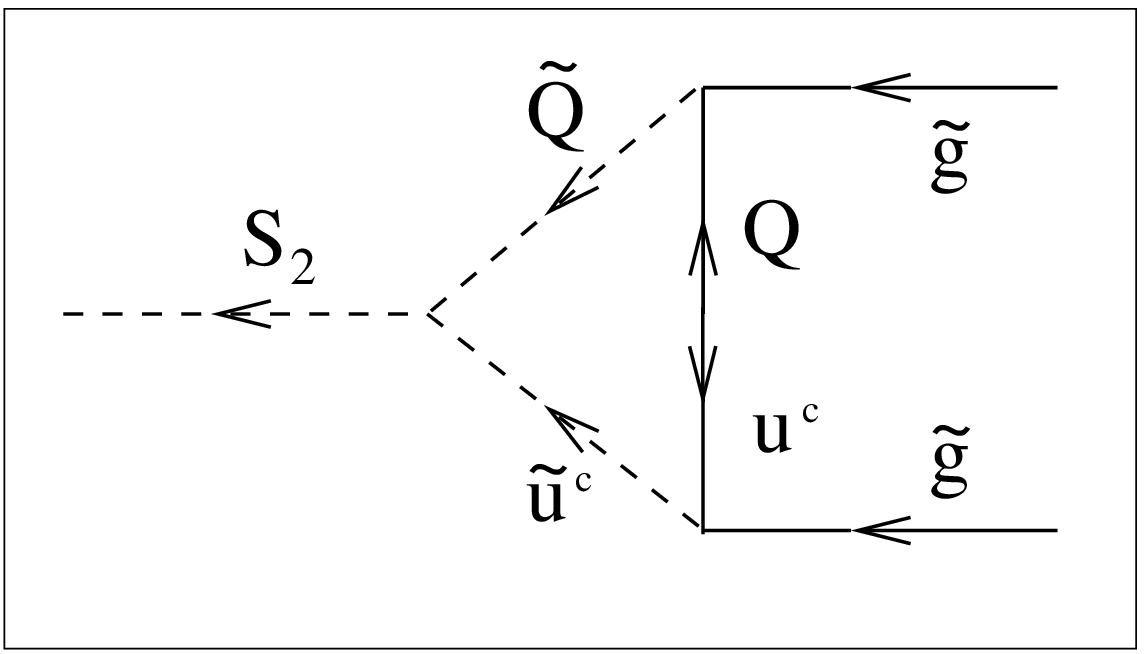}}}
\caption{ The diagrams that generate the triscalar coupling $A_u$ (left), and 
the gluino mass (right).} 
\end{figure}                

Using Eq.~(18), we also obtain the effective interaction (see Fig.~1, right 
window) given by
\begin{equation}
{g_s^2 \over 16 \pi^2} h_2^* h_u {v_d m_u \over M^2_{eff}} S_2^* \tilde g 
\tilde g,
\end{equation}
which generates a gluino mass proportional to $v_2$, together with a 
dynamical phase from $S_2$.  This solves the strong CP problem as 
proposed in Ref.\cite{dema}.  However, the generated mass itself is very 
small.  Even with $m_t = 174$ GeV, $M_{\tilde g}$ is at most a few GeV. 
On the other hand, such a light gluino is not completely ruled out 
experimentally and may yet be discovered \cite{light}.  To obtain a heavy 
gluino (with mass greater than 250 GeV), some new physics at the TeV scale 
will be required.  For example, consider the addition of a neutral singlet 
$\hat \chi$ with $R=0$ and colored triplets $\hat \psi$ and $\hat \psi^c$ 
with $R=1$.  All are assumed to transform as $\omega^2$ under $Z_3$.  Then 
the extra terms in the superpotential, i.e.
\begin{equation}
f_2 \hat S_2 \hat \chi \hat \chi + f_0 \hat \chi \hat \psi \hat \psi^c,
\end{equation}
will generate a gluino mass given by
\begin{equation}
M_{\tilde g} = {g_s^2 \over 8 \pi^2} f_2 v_2,
\end{equation}
where the masses of $\tilde \psi$, $\tilde \psi^c$ and the Dirac fermion 
formed out of $\psi$ and $\psi^c$ are all taken to be $f_0 \langle \chi 
\rangle$.  In Eq.~(15), let $|\mu_1| = 4$ TeV, $\mu_{12} = 0.4$ TeV, then 
$v_2 = 10$ TeV.  Now let $f_2 = 1.3$, then $M_{\tilde g} = 250$ GeV.

The usual incorporation of the axion into a supersymmetric model assumes 
the original Peccei-Quinn symmetry \cite{pq} for the corresponding 
superfields, i.e. +1/2 for $\hat Q$, $\hat u^c$, $\hat d^c$, $\hat L$, 
$\hat e^c$, and $-1$ for $\hat H_u$, $\hat H_d$.  This assignment forbids 
the $\mu$ term in the MSSM superpotential, as well as the $B$ term in 
$V_{soft}$.  Assuming supergravity, the $U(1)_{PQ}$ symmetry is then broken 
together with local supersymmetry in the K\"{a}hler potential at the axion 
scale $f_a$.  The effective scale of global supersymmetry breaking becomes 
of order $f_a^2/M_{Planck}$.  The singlet superfield carrying the axion is 
here some kind of ``messenger'' field which communicates between the MSSM 
and the hidden sector.  In our case, the axion comes from a superfield 
which lives entirely in our world.  The origin of soft supersymmetry breaking 
at the TeV scale is not specified, only that it has to preserve $U(1)_R$.  
Given such a structure and with the help of a $Z_3$ discrete symmetry which 
is softly broken also at the TeV scale, we find that the $U(1)_R$ symmetry 
is actually broken spontaneously at a scale much larger than $M_{SUSY}$. 
In fact, this mechanism also works if we use the original $U(1)_{PQ}$ 
instead of $U(1)_R$.  In that case, $\hat S_{2,1,0}$ should have $PQ$ 
charges of +2, -1, and $-2$ respectively.  Hence the superpotential of Eq.~(8) 
is obtained without using the $Z_3$ discrete symmetry \cite{future}.

In conclusion, we have succeeded in formulating a realistic supersymmetric 
model with a spontaneously broken $U(1)_R$ symmetry as the natural solution 
of the strong CP problem.  The soft breaking of the supersymmetry at the 
TeV scale induces an axion scale of order $\sqrt {m_2 M_{SUSY}}$, where 
$m_2$ is some unification scale, such as the string scale or the Planck 
scale.  We have thus a first example of the unusual situation where 
the mass of the physical field ($S_1$) is much smaller than its VEV.
This is in contrast to the less uncommon occurrence \cite{lima} where 
the mass of the physical field ($S_2, S_0$) is much greater than its VEV. 

The resulting model resembles closely the MSSM, with the following 
distinctions.  (1) The $\mu$ parameter is replaced by $h_2 \hat S_2$ in 
the superpotential, thus solving the so-called $\mu$ problem.  (2) Although 
the breaking of $U(1)_R$ by $v_1$ and $v_2$ also breaks $R$ parity, the 
latter is effectively conserved as far as the MSSM particles are concerned 
because its violation is suppressed by $v_2/v_1$.  (3) Supersymmetric scalar 
masses and the equivalent $B$ term are as in the MSSM.  However, $U(1)_R$ 
invariance forbids $A$ terms and gaugino masses.  (4) The spontaneous 
breaking of $U(1)_R$ leads to $A$ terms proportional to $v_2$ and to the 
corresponding Yukawa coupling matrix, thereby suppressing flavor-changing 
neutral currents automatically.  (5) Gaugino masses are generated 
radiatively but are probably too small to be realistic.  They can be made 
larger by the addition of new particles at the TeV scale.  (6) In contrast 
to the DFSZ and KSVZ models, the new particles associated with the axion, 
i.e. those of Eq.~(11), are now at the TeV scale.  However, their couplings 
with the MSSM particles are all suppressed by $v_2/v_1$, so they are 
effectively unobservable at future colliders \cite{marasa}.\\[5pt]

This work was supported in part by the U.~S.~Department of Energy
under Grants No.~DE-FG03-94ER40837 (E.M.) and No.~DE-FG02-94ER40823 (D.A.D.)

\newpage
\bibliographystyle{unsrt}

\begin{thebibliography}{99}
\bibitem{future} Details will be given in a forthcoming paper comparing the 
phenomenological consequences of the two different approaches.
\end{thebibliography}

\end{document}